\documentclass [journal]{IEEEtran}
\usepackage{amssymb}
\usepackage{cite}

\ifCLASSINFOpdf
\else
\fi
\usepackage{amsfonts}
\usepackage{amsmath}
\usepackage{latexsym,bm}
\usepackage{amsthm}

\usepackage{array}
\usepackage{blindtext}
\usepackage{fixltx2e}
\usepackage{graphicx}
\usepackage{epstopdf}
\usepackage{float}
\usepackage{subfigure}
\usepackage{algorithmicx}
\usepackage{algorithm}
\usepackage{algpseudocode}

\usepackage{color}
\usepackage{multirow}
\usepackage{mathrsfs}
\begin{document}
%
\title{Handover Control in Wireless Systems via Asynchronous Multi-User Deep Reinforcement Learning}
%
%
%

%
%

\author{Zhi~Wang, Lihua~Li,~\IEEEmembership{Member, IEEE}, Yue~Xu, Hui~Tian,~\IEEEmembership{Member, IEEE}, and Shuguang~Cui,~\IEEEmembership{Fellow, IEEE}
\IEEEcompsocitemizethanks{Z. Wang, L. Li and H. Tian are with the State Key Laboratory of Networking and Switching Technology,~Beijing University of Posts and Telecommunications, Beijing, China.}
\IEEEcompsocitemizethanks{Z. Wang, Y. Xu and S. Cui are with the Department of Electrical and Computer Engineer, University of California at Davis, CA 95616, USA.}
\IEEEcompsocitemizethanks{Y. Xu is with the Key Laboratory of Universal Wireless Communications, Ministry of Education
Beijing University of Posts and Telecommunications, Beijing, China.}
\IEEEcompsocitemizethanks{Corresponding author: S. Cui (e-mail:sgcui@ucdavis.edu).
}}


%



\maketitle
\begin{abstract}
In this paper, we propose a two-layer framework to learn the optimal handover (HO) controllers in possibly large-scale wireless systems supporting mobile Internet-of-Things (IoT) users or traditional cellular users, where the user mobility patterns could be heterogeneous. In particular, our proposed framework first partitions the user equipments (UEs) with different mobility patterns into clusters, where the mobility patterns are similar in the same cluster. Then, within each cluster, an asynchronous multi-user deep reinforcement learning scheme is developed to control the HO processes across the UEs in each cluster, in the goal of lowering the HO rate while ensuring certain system throughput. In this scheme, we use a deep neural network (DNN) as an HO controller learned by each UE via reinforcement learning in a collaborative fashion.  Moreover, we use supervised learning in initializing the DNN controller before the execution of reinforcement learning to exploit what we already know with traditional HO schemes and to mitigate the negative effects of random exploration at the initial stage. Furthermore, we show that the adopted global-parameter-based asynchronous framework enables us to train faster with more UEs, which could nicely address the scalability issue to support large systems. Finally, simulation results demonstrate that the proposed framework can achieve better performance than the state-of-art on-line schemes, in terms of HO rates.
\end{abstract}

\section{Introduction}
Global mobile data traffic has recently seen a significant increase. In order to address this challenge, the ultra-dense network (UDN) is being considered as one of the potential solutions, which is realized by deploying small base stations (SBSs) densely at the traffic hotspots. In cellular networks, to maintain the service, the user equipment (UE) changes its serving base station (BS) as it moves, which is called handover (HO). Traditionally, the HO process is designed for macro-cell systems and triggered by the event A3 of UE \cite{RRC}, when the difference between the periodically measured Reference Signal Received Power (RSRP)/ Reference Signal Received Quality (RSRQ) of a candidate cell (target cell) and that of the serving cell is higher than the HO hysteresis margin (HHM). Once A3 is triggered, the UE will wait for a predetermined time duration, i.e., time-to-trigger (TTT); afterwards, if the triggering condition keeps being satisfied, the HO is finalized and the UE is connected to the target cell. In the traditional macro-cell system, static HHM and TTT strategies are usually adoptable.

However, in UDNs, the HO strategy for macro-cell systems may lead to the frequent HO problem, where the HO process can be triggered even with a little movement of the UE \cite{7572176}. Frequent HOs may diminish the capacity gain obtained by network densification as the HO processes break the data flows while terminating the serving link and reestablishing the link to the target BS, which is called the HO delay \cite{7792669}. Moreover, the increased HO overheads from the serving BSs, target BSs, UEs, and core networks consume more energy and other resources. Hence, to utilize the full potential of UDN, it is essential to maintain an appropriate HO rate, defined as the expected number of HOs per unit time. In \cite{6477064,7110473}, the performance analysis over HO rates was conducted for the irregularly shaped network topologies, where no solutions were proposed for the frequent HO problem.


One way to optimize the HO process is tuning the HO parameters adaptively by implementing threshold comparisons with several specific metrics \cite{6067552,5629008,7341220}. In \cite{6067552}, an adaptive HHM approach was proposed to lower the number of HOs, which uses a predefined RSRQ threshold and path loss factor to adapt the HHM. In  \cite{5629008}, the authors defined a weighted-sum cost function, which consists of key parameters related to the UE speed, cell load, and number of user connections. Then, the cost function was integrated into a typical RSRP-based procedure as an additive factor to change HHM adaptively. In  \cite{7341220}, a fuzzy logic controller that can tune the HHM adaptively was proposed to decrease the signaling load caused by HO. However, the above threshold-comparison based HO-tuning methods lack systematic methodologies to optimize the algorithm-related parameters, e.g., the number of rule bases \cite{7341220}, which may not be acceptable in practical applications.

Another type of HO optimization strategies is based on dynamic programming (DP). Specifically, the HO decision process is modeled as a Markov Decision Process (MDP) \cite{712192} and the objective is to obtain the optimal policy, which is the probability distribution of actions (HO decisions) conditioned on the input state. In \cite{7725961}, an upper bound for the HO performance in UDNs was derived, where the Viterbi algorithm was used to derive the optimal HO decision policy under the ideal assumption that the positions of SBSs or the UE trajectories are available in advance. In \cite{6777395}, one HO decision policy was proposed to minimize the HO rate while maintaining certain system throughput. Specifically, with prior knowledge of the cell load transition probabilities, the state observations were first converted into belief states. Then the optimal policy to maximize the expected reward over the belief state was derived. In \cite{7314972}, a TTT selection policy was proposed by assuming the knowledge of the UE trajectory distribution and channel propagation model. All the above results \cite{7725961,6777395,7314972} adopted a similar methodology, which is to first compute certain transit probabilities with the knowledge of some network dynamics, e.g., UE trajectories \cite{7314972} or load state transition probabilities \cite{6777395}, and then use the derived transit probabilities to derive the optimal HO decision policy. In other words, the above methods to obtain the HO decision policy are falling into the traditional model-based framework. However, in practice, it is difficult to assume strong prior knowledge on network dynamics as required in the model-based methods. Hence, it is highly desirable that the method to derive the optimal HO decision policy does not depend on any prior knowledge of network dynamics, i.e., we need to seek model-free or data-driven approaches.

The most similar work to our proposed method is given in \cite{7572176,7925950}, where the authors proposed some model-free learning-based approaches to solve the frequent HO problem. In particular, the upper confidence bandit (UCB) algorithm was used to derive the optimal HO policy, which minimizes the number of HO occurrences while ensuring certain throughput, outperforming the 3GPP protocol by up to $80\%$. Moreover, the UCB algorithm samples rewards from the reward distribution of SBSs continually and tries to find the single best SBS with the largest mean reward to camp on consistently.

However, two major issues exist in applying the method mentioned above to the considered mobility optimization problem. First, in large-scale systems including different scenarios, e.g., offices vs. malls, the mobility patterns may vary dramatically across different areas, which needs different optimal controllers. Hence, when the UEs enter some new scenarios, they need to learn from random starting points under the setup in \cite{7572176,7925950}, which may degrades the performance; second, for newly arriving UEs in one scenario, the learning has to start from some possibly ill-performed initial points, such that the UEs may suffer from unacceptable performance, possibly over a long period.

Here, we propose a two-layer framework to address the above issues. Specifically, we first adopt a centralized controller that partitions UEs into clusters with unsupervised learning \cite{WARRENLIAO20051857}, where the UEs in different areas can have different mobility patterns. After the clustering, the HO processes of UEs in the same cluster are modeled as the similar MDPs. We adopt the reinforcement learning (RL) framework to learn the optimal controller for each UE, which makes HO decisions. We incorporate the situation and exploration information of UEs into states to model the HO process, where the exploring information, i.e., the serving SBS indices for UEs, can facilitate the exploitation versus exploration tradeoff to accelerate the learning. However, such an adopted state space could be large, which is hard to track. Recently, deep neural network (DNN) \cite{mnih-dqn-2015} was proposed to approximate the values of $Q-$functions\footnote{The action value function, $Q-$function, is used to generate appropriate policies, which will be introduced later. } for all the states, which is called DQN.  To make the DQN more suitable to RL\footnote{The DQN works well when the data distribution is stationary, which is not the case in most RL applications. In particular, the policy improvement process, which will be introduced later, can dramatically change the state occurrence probability. Moreover, the consecutive states are correlated strongly in RL systems.}, the authors in \cite{mnih-dqn-2015} adopted the techniques of experience replay and delayed updated target networks to stabilize the learning and prevent divergence of RL algorithms. Many approaches were later proposed to improve the stability, convergence, and learning speed of DQN, e.g., the dueling network structure \cite{Wang2016} and the fast learning technique \cite{He2017}.

In this paper, our contributions are summarized as follows.
 \begin{itemize}
 \item Firstly, we propose the two-layer framework to learn optimal HO controllers in large-scale systems with different scenarios. In particular, the centralized controller can cluster the UEs according to their mobility patterns with unsupervised learning, where the mobility patterns are similar in the same cluster. We then use the RL framework to obtain the optimal controller for each UE with the same cluster.
 \item Secondly, we adopt a model-free asynchronous advantage actor-critic (A3C) \cite{pmlr-v48-mniha16} RL framework to achieve the optimal HO policy in each cluster. Note that such an asynchronous framework in A3C can accelerate learning when the number of UEs increases, while the learning time increases at least linearly against the number of UEs with the introduced methods in \cite{7572176,7925950}. Also, we propose to use certain situation and exploration information as the states to derive better performance.
 \item Thirdly, we utilize DNN to approximate the $Q-$function and generate the policy in A3C. Due to the generalization ability of DNN \cite{712192}, such function approximators can represent the whole state space with limited weights, which can avoid the degraded performance for the clustered newly-arriving UEs during the learning transitions. Specifically, after the clustering for the newly-arriving UEs, the clustered newly-arriving UEs fetch the DNN weights from global parameter servers of their clusters as the pretrained networks, which are better than random initialized networks. We then propose two methods to utilize this pretained networks: on-line vs. off-line. In the on-line method, the UEs keep learning and fetch the weights from the parameter server as the pretrained network periodically. While in the off-line method, the UEs treat the pretrained networks as static controllers.\footnote{It is worth noting that if the DNN cannot generalize the whole state space, it may lead to degraded performance for the UEs using the pretrained network.}
 \item  Finally, to improve the performance at the very beginning for both the on-line and off-line methods, i.e., at the time before the RL algorithm is executed to learn the DNN, we initialize the DNN with supervised learning (SL), where the training data set is from the output of the traditional HO scheme proposed in 3GPP.
\end{itemize}

The rest of the paper is organized as follows. The preliminaries about MDP and RL are presented in Section II. The system model is introduced in Section III, with the clustering process in section III-A, the asynchronous joint learning framework in Section III-B, the state vector design in III-C, and the reward signal design in Section III-D. In Section IV, the policy optimization problem is formulated in Section IV-A; the learning algorithm is discussed in Section IV-B; the algorithm implementation framework is explained in Section IV-C; the DQN structure and network initialization with supervised learning are presented in Section IV-D and Section IV-E, respectively. Simulation setups and results are provided in Section V. Finally, Section VI concludes the paper.

\section{Preliminary}
Before proceeding to describe the system model, we first describe the necessary frameworks, i.e., the Markov decision process (MDP) and the general RL algorithm frameworks, where the value function and policy gradient methods are introduced, which will be utilized in our learning scheme. Finally, a special kind of recurrent neural network (RNN), i.e., the long-short term memory (LSTM) network, is presented, which will be utilized as the DQN function approximator in the proposed RL framework.

\subsection{Markov Decision Process}
\begin{figure}[!htbp]
\setlength{\abovecaptionskip}{0pt}
\centering
\includegraphics [width=\linewidth]{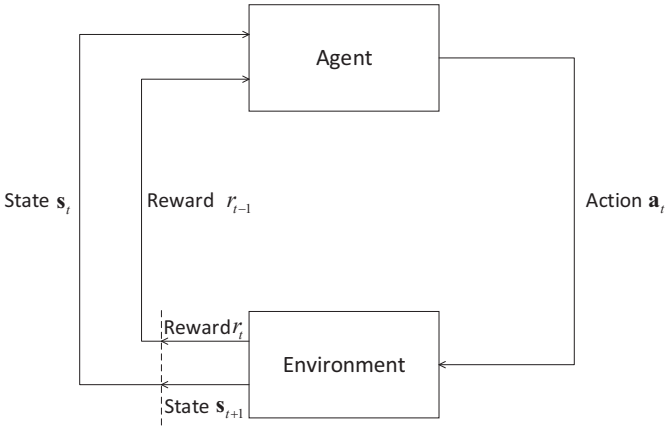}
\caption{ General framework of RL algorithm. }
\label{fig1}
\end{figure}
Typically, the RL problem is modeled as a MDP, i.e.,  $\left\langle {{\bf{S}},{\bf{A}},{\bf{P}},{\bf{\pi}}, r,\gamma } \right\rangle $, which is composed of a finite state space ${\bf{S}}$, an action space ${\bf{A}}$, a transit probability set ${\bf{P}}$ mapping each point $({\bf{s}},{\bf{a}}) \in {\bf{S}} \times {\bf{A}}$ to the next state ${\bf{s}}' \in {\bf{S}}$, a policy ${\bf{\pi}}:{\bf{S}} \to {\bf{A}} $, a reward function $r: {\bf{S}} \times {\bf{A}} \to \mathcal{R}$, and a discount factor $\gamma$ penalizing the future rewards. Specifically, at time step $t$, the agent senses the state ${\bf{s}}_{t}$ and takes action ${\bf{a}}_{t}$ according to policy $\pi({\bf{a}}_{t}|{\bf{s}}_{t})$. Then the environment feeds back the reward signal $r_t$ and the new state ${\bf{s}}_{t+1}$ to the agent. Hence, the general framework of RL is shown in Fig. \ref{fig1}. The goal of the agent is to obtain a policy maximizing the average cumulative reward in the long run. In a typical RL framework, two value functions are defined as:
\begin{align}
&Q^{\pi}({\bf{s}}_{t},{\bf{a}}_{t})=E_{\pi}\left(\sum\limits_{k = 0}^{T-t-1} {\gamma^k {r_{t+k}}} | {\bf{s}}_{t},{\bf{a}}_{t} \right)= \sum\limits_{k = 0}^{T-t-1} {{\gamma ^k}{{({P^{\pi} })}^k}} {r_{t}}\label{1},\\
&V^{\pi}({\bf{s}}_{t})=E_{\pi}\left(\sum\limits_{k = 0}^{T-t-1} {\gamma^k {r_{t+k}}} | {\bf{s}}_{t}\right)=\sum\limits_{{\bf{a}}_{t}}\pi({\bf{a}}_{t}|{\bf{s}}_{t}) Q^{\pi}({\bf{s}}_{t},{\bf{a}}_{t})\label{2}
\end{align}
with $P^{\pi}$ defined as an operator over reward $r_{t}$, i.e.,
\begin{equation}
({{P^{\pi} }})r_{t} \buildrel \Delta\over = \sum\limits_{{\bf{s}}_{t+1},{\bf{a}}_{t+1}} {P({\bf{s}}_{t+1}|{\bf{s}}_{t},{\bf{a}}_{t})\pi ({\bf{a}}_{t+1}|{\bf{s}}_{t+1})r_{t+1}},
\label{21}
\end{equation}
where $(X)^k$ denotes $k$ successive applications of operator $X$, $P({\bf{s}}_{t+1}|{\bf{s}}_{t},{\bf{a}}_{t})$ is the transmit probability and $T$ is the terminal time step\footnote{If $T<\infty$, it is a episodic task, which means that the agent-environment interaction breaks naturally into episodes and each episode ends in a special state called the terminal state. Otherwise, it is a continuing task without a terminal state. }. Moreover, $Q^{\pi}({\bf{s}}_{t},{\bf{a}}_{t})$ and $V^{\pi}({\bf{s}}_{t})$ are called action-value and state-value functions, respectively. We see that the action-value function $Q^{\pi}({\bf{s}}_{t},{\bf{a}}_{t})$ (or state-value function $V^{\pi}({\bf{s}}_{t})$) describes the expected discounted sum of rewards starting from state-action pair $({\bf{s}}_{t},{\bf{a}}_{t})$ (or state ${\bf{s}}_{t}$) over a given policy $\pi$. Hence, the goal of an RL algorithm is to obtain the optimal policy $\pi^*$ by solving following problem:
\begin{equation}
\pi^* = \arg \max\limits_\pi V^\pi({\bf{s}}_{t})\label{3}
\end{equation}
Furthermore, we define two operators as
\begin{align}
&\left( {{\Gamma ^\pi }} \right)Q({\bf{s}}_{t},{\bf{a}}_{t}) \buildrel \Delta \over = r_t + \gamma\left( {{P^\pi }} \right)Q({\bf{s}}_{t},{\bf{a}}_{t})\\
&\left( {{\Gamma ^*}} \right)Q({\bf{s}}_{t},{\bf{a}}_{t}) \buildrel \Delta \over = r_t + \gamma\max\limits_{\pi} \left( {{P^\pi }} \right)Q({\bf{s}}_{t},{\bf{a}}_t)
\end{align}
which are called the Bellman operator and the optimal Bellman operator, respectively. Moreover, $\left( {{P^\pi }} \right)Q({\bf{s}}_{t},{\bf{a}}_{t})= \sum\nolimits_{{\bf{s}}_{t+1},{\bf{a}}_{t+1}} P({\bf{s}}_{t+1}|{\bf{s}}_{t},{\bf{a}}_{t})\pi ({\bf{a}}_{t+1}|{\bf{s}}_{t+1})Q({\bf{s}}_{t+1},{\bf{a}}_{t+1})$. Note that $Q^\pi$ is the unique fixed point of operator $\Gamma^\pi$, i.e., $\Gamma^\pi Q^\pi=Q^\pi$, and $Q^*$ is the optimal action-value function under $\pi^*$ with $\Gamma^* Q^*=Q^*$. Moreover, $V^\pi$ and $V^*$ are also the fixed points of $\Gamma^\pi$ and $\Gamma^*$, respectively.
\subsection{Value-Function Prediction with Deep Neural Network }
Generally, value-function prediction is to derive the value functions with a given policy. The intuitive way to calculate the value functions is based on the definitions in (\ref{1}) and (\ref{2}). However, this is usually not feasible as the transit probability is difficult to obtain in practical systems. In \cite{712192}, function approximators were proposed to estimate the value functions. For this purpose, DNN approximators could be used to generalize the whole state space using limited parameters \cite{mnih-dqn-2015,pmlr-v48-mniha16}, when the state space is large. Hence, the prediction problem for the action-value function is equivalent to solving the  following the optimization problem:
\begin{equation}
\min\limits_{\bf{w}} J({\bf{w}}) = \sum\limits_{{\bf{s}}_{t}\in{\bf{S}} } \mu({\bf{s}}_{t})\sum\limits_{{\bf{a}}_{t}\in{\bf{A}} } \pi({\bf{a}}_{t}|{\bf{s}}_{t})(Q^\pi({\bf{s}}_{t},{\bf{a}}_{t})-Q_{\bf{w}}({\bf{s}}_{t},{\bf{a}}_{t}))^2,
\end{equation}
where ${\bf{w}}$ is the approximator parameter (e.g., the DNN weights) to be determined and $\mu({\bf{s}}_{t})$ is the state distribution, which depends on transit probability and policy, thus unknown in general. To solve the above optimization problem, algorithms based on stochastic gradient decent (SGD) could be adopted. Specifically, at every time step $t$, the parameter ${\bf{w}}$ is updated as:
\begin{equation}
{\bf{w}}  \leftarrow {\bf{w}} + \alpha [Q^\pi({\bf{s}}_{t},{\bf{a}}_{t})-Q_{\bf{w}}({\bf{s}}_{t},{\bf{a}}_{t})] \nabla_{\bf{w}} Q_{\bf{w}}({\bf{s}}_{t},{\bf{a}}_{t})\label{4},
\end{equation}
where $\alpha$ is a positive step-size and the state-action pair $({\bf{s}}_{t},{\bf{a}}_{t})$ is sampled from the environment following a given policy. However, the exact update in (\ref{4}) cannot be performed since the update target $Q^\pi({\bf{s}}_{t},{\bf{a}}_{t})$ is unknown. Therefore, the update rule could be modified as
\begin{equation}
{\bf{w}} \leftarrow {\bf{w}} + \alpha [G_t-Q_{\bf{w}}({\bf{s}}_{t},{\bf{a}}_{t})] \nabla_{\bf{w}} Q_{\bf{w}}({\bf{s}}_{t},{\bf{a}}_{t}),
\end{equation}
where $G_t$ is an estimated target approximating the true value $Q^\pi({\bf{s}}_{t},{\bf{a}}_{t})$. Typically, the target $G_t$ has two main choices: the Monte-Carlo vs. bootstrapping target. In particular, the Monte-Carlo target is given as $G_t  \buildrel\textstyle.\over= r_t + \gamma r_{t+1} + ...\gamma^{T-t-1} r_T$ and a TD(0) \cite{712192} bootstrapping target is given as $G_t\buildrel\textstyle.\over=r_t + \gamma Q_{\bf{w}}({\bf{s}}_{t+1},{\bf{a}}_{t+1}) $.

\subsection{Policy Gradient Methods}
In the above, the value-function prediction methods estimate the value function first, based on which the policy improvement is performed. Policy gradient methods instead parameterize the policy with ${\bf{\theta}}$ and use SGD to update ${\bf{\theta}}$, thus the policy directly. In particular, the gradients over ${\bf{\theta}}$ is obtained from a derivation of the objective function in (\ref{3}) \cite{sutton2000policy}:
\begin{align}
\nabla_{{\bm{\theta}}} V^{\pi_{{\bm{\theta}}}}({\bf{s}}_{t})
=E_{\pi_{{\bm{\theta}}}}\left[{{\nabla _{{\bm{\theta}}} }\log {\pi _{{\bm{\theta}}} }({{\bf{a}}_{t}}|{\bf{s}}_{t})}Q^{\pi_{{\bm{\theta}}}}({\bf{s}}_{t},{\bf{a}}_{t}) \right].
\label{7}
\end{align}
Hence, at time step $t$, the gradient used to update parameter ${\bm{\theta}}$ is sampled from (\ref{7}) (thus stochastic gradient), given as
\begin{equation}
\Delta {\theta}={{\nabla _{{\bm{\theta}}} }\log {\pi _{{\bm{\theta}}} }({{\bf{a}}_{t}}|{\bf{s}}_{t})}Q^{\pi_{{\bm{\theta}}}}({\bf{s}}_{t},{\bf{a}}_{t}).
\label{20}
\end{equation}
To compute the terms in (\ref{20}), we could use the previously introduced value-function prediction methods to estimate $Q^{\pi_{{\bm{\theta}}}}({\bf{s}}_{t},{\bf{a}}_{t})$.

\begin{figure}[!htbp]
\setlength{\abovecaptionskip}{0pt}
\centering
\includegraphics [width=\linewidth]{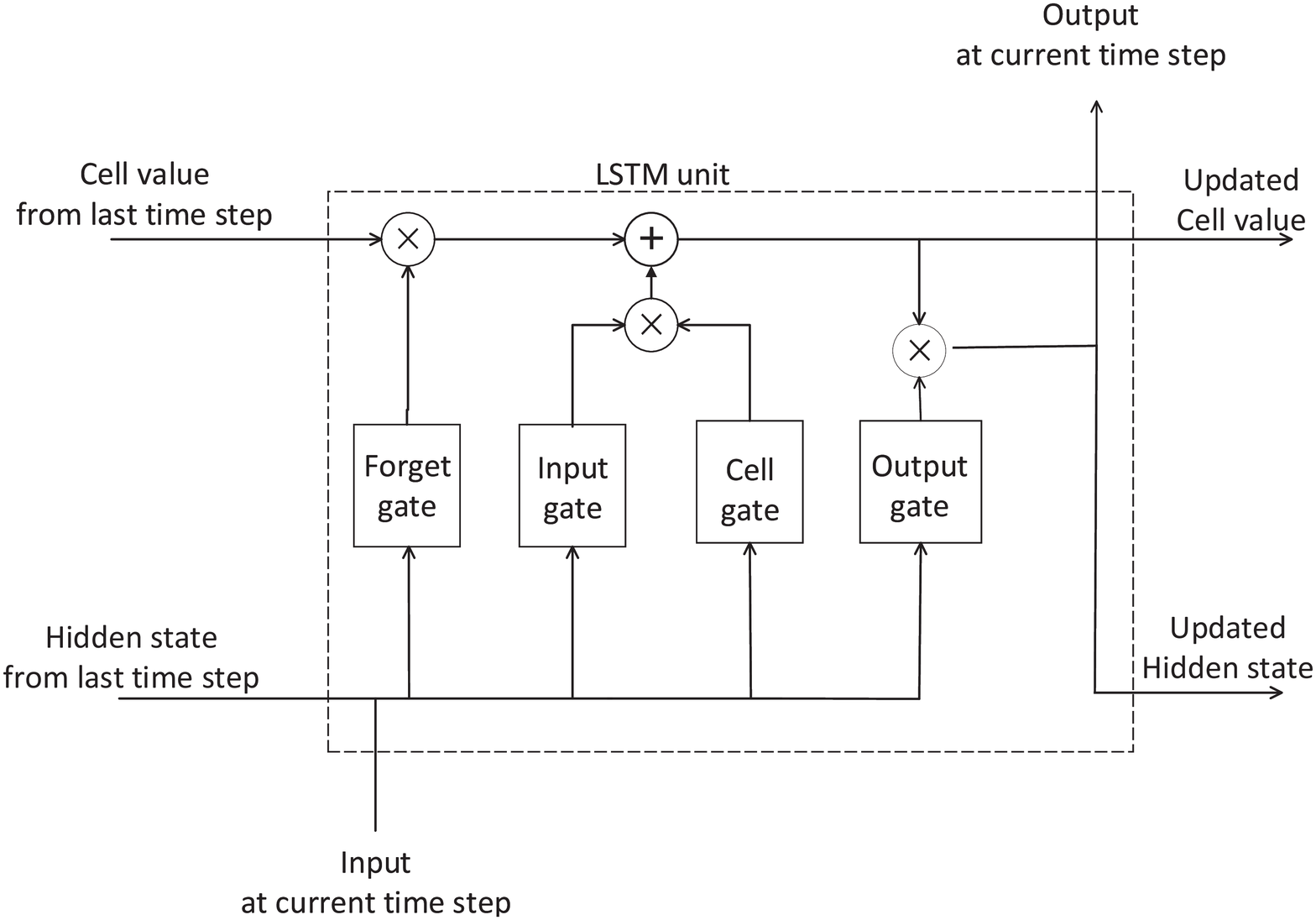}
\caption{ Structure of LSTM unit. }
\label{fig11}
\end{figure}
\subsection{LSTM}
An LSTM network is capable of learning long-term dependencies. In particular, an LSTM network is considered as multiple copies of the LSTM unit, each of which passes a message to its successor. Moreover, as shown in Fig. \ref{fig11}, an LSTM unit utilizes four gates to keep long-term information from previous time steps. The information transferred between the time steps include the cell state and hidden state. The first operation in an LSTM unit is to decide what information to throw away from the cell state, which is done by multiplying the output of forgotten gate by the cell state from the last time step. The next operation is to store new information in the cell state, which has two parts: the cell gate first generates the candidate cell state value; and the input gate then weighs the candidate cell state value. The weighted candidate cell state is then added to the multiplied cell state by the forgotten gate, to update the cell state. Finally, the hidden state is decided by the updated cell state and the output of the output gate, which is considered as the output of the LSTM unit for forward-propagation and back-propagation. In our DQN approximator, we add a fully connected neuron layer to generate the input for each LSTM unit. In addition, the output of the LSTM unit is processed by two separated neuron output layers to generate the value function and the policy, respectively, as later shown in Fig. \ref{fig5}.

\section{System Model}

In the system, we consider $L$ areas, denoted as $\{\mathscr{A}_l\}, l \in \{1,2,...,L\}$, which could be nonadjacent physically and represent different scenarios, e.g., offices vs. malls. Furthermore, $K_l$ SBSs exist and $N_l$ UEs move around in area $\mathscr{A}_l$. Hence, we denote the set of all the UEs as ${\bf{Z}}=\{{\rm{UE}}_{i,l}\}$, $i\in{1,..,N_l}$, $l\in{1,..L}$.  The centralized controller partitions the UEs in the $L$ areas into $H$ clusters according to the user mobility patterns, i.e., cluster ${{\bf{G}}_h} = \{{\rm{UE}}_{i,l,h}\}\subseteq{\bf{Z}}, h\in\{1,..,H\}$ and ${\rm{UE}}_{i,l,h}$ represents the UE $i$ in area $l$ physically partitioned into cluster $h$ logically. In addition, we have ${\bf{Z}} = \bigcup\nolimits_{h = 1}^H {{{\bf{G}}_h}}$ and ${{\bf{G}}_{{h_1}}}\bigcap {{{\bf{G}}_{{h_2}}} = \emptyset }$, for $h_1  \ne h_2 $. In this paper, for the convenience of analysis, we assume that the UEs share similar mobility patterns in the same area. Hence, it is reasonable to assume that the UEs in the same area will be clustered into the same cluster.  At time step $t$, the $i-$th UE in area $l$ and cluster $h$ maintains its active SBS index set by a vector ${\bf{b}}_{i,l,h,t}, i \in \{1,2,...,N_l\}$, $l\in{1,..L}$ with $|{\bf{b}}_{i,l,h,t}| = M_l$. For simplicity, we assume that $M_l=K_l$, and the system operates over equal-length time slots and the HO decision of UE $i$ is made at the beginning of each time slot, where UE $i$ chooses a SBS from ${\bf{b}}_{i,l,h,t}$ to camp on. If the UE is out of the system coverage at the beginning of time slot $t$, the current state is noted as the terminal state and time $t$ thus becomes the terminal time step $T$, i.e., $T=t$, which ends the current episode. The HO controllers are learned in each cluster using an asynchronous learning framework, which will be discussed in details later. We model the HO process of UE $i$ in area $l$ and cluster $h$ as a discrete-time episodic MDP process $\left\langle {{\bf{S}}_{i,l,h},{\bf{A}}_{i,l,h},{\bf{P}}_{i,l,h},{\bf{\pi}}_{i,l,h}, r_{i,l,h},\gamma } \right\rangle $.
\begin{figure}[!htbp]
\setlength{\abovecaptionskip}{0pt}
\centering
\includegraphics [width=\linewidth]{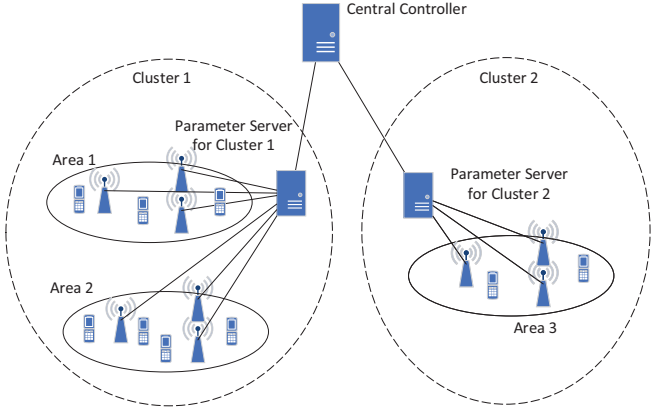}
\caption{ An example of our proposed framework is represented, where three areas, i.e., $\mathscr{A}_1$, $\mathscr{A}_2$ and $\mathscr{A}_3$ are not adjacent. We assume that the mobility patterns are similar in $\mathscr{A}_1$ and $\mathscr{A}_2$ and different from that in $\mathscr{A}_3$. Hence, the centralized controller can partition the UEs in the three areas into two clusters according to the mobility patterns. The parameter server in the same cluster enables parameter sharing among the UEs. }
\label{fig2}
\end{figure}

In Fig. \ref{fig2}, an example of our proposed two-layer framework with three areas is presented. To clarify this framework, in the following, we first introduce the centralized controller and then the asynchronous learning framework within each cluster. Afterwards, the designs of the state vector and the reward signal will be discussed.
\subsection{Clustering by Centralized Controller}

As mentioned above, the centralized controller partitions the UEs into clusters based on their mobility patterns, which are then assumed similar across UEs in the same cluster. According to \cite{7904647}, the mobility patterns can be extracted from geographic contexts, which include locations of the UEs, often enriched with speed information as well as past trajectories. However, these geographic contexts typically cannot be obtained directly in cellular networks, for which some mobility tracking algorithms have been proposed to obtain the locations and the speeds of the UEs from the RSRP \cite{1390891}. In this paper, we assume that we have already obtained the geographical contexts after implementing certain mobility tracking algorithm. Specifically, the feature vector ${\bf{d}}_{i,l}$ of UE $i$ in area $l$ is defined as
\begin{equation}
{\bf{d}}_{i,l} = \{{\bf{d}}_{i,l}^1,...,{\bf{d}}_{i,l}^{T_u}\},
\end{equation}
where $T_u$ is a fixed time period for mobility observations and ${\bf{d}}_{i,l}^t, t\in\{1,...T_u\}$ contains the 2-dimensional coordinates and speeds, i.e., ${\bf{d}}_{i,l}^t =\{x_{i,l}^{t},y_{i,l}^{t},v_{i,l}^t\}$. Therefore, the input data set to the centralized controller is ${\bf{D}} = \{{\bf{d}}_{i,l}\}, i\in\{1,..,N_l\}, l\in\{1,...,L\}$.

We then utilize a standard K-means clustering algorithm \cite{macqueen1967} to partition UEs into $H$ clusters with the input data set ${\bf{D}}$. Specifically, the objective of K-Means clustering is to minimize the total intra-cluster variance:
\begin{equation}
\min\limits_{\{{\bf{o}}_h\}} \sum\limits_{h=1}^H \sum\limits_{i=1}^{N_l}\sum\limits_{l=1}^{L} dis({\bf{d}}_{i,l}, {\bf{o}}_h),
\end{equation}
where ${\bf{o}}_h=\{{\bf{o}}_h^t\}, t \in \{1,...,T_u\} $ is the centroid of cluster $h$, ${\bf{o}}_h^t$ has the same length as ${\bf{d}}_{i,l}^t$, i.e., ${\bf{o}}_h^t=\{{o_{h,x}^t,o_{h,y}^t,o_{h,v}^t}\}$, the distance between ${\bf{d}}_{i,l}$ and ${\bf{o}}_h$ is defined as $dis({\bf{d}}_{i,l}, {\bf{o}}_h)=\tau[\sum\nolimits_{t=1}^{T_u}(x_{i,l}^t-o_{h,x}^t)^2+(y_{i,l}^t-o_{h,y}^t)^2]+(1-\tau)\sum\nolimits_{t=1}^{T_u}(v_{i,l}^t-o_{h,v}^t)^2$, and $\tau$ is a weight factor. The full algorithm is given in Algorithm \ref{alg}.
\begin{algorithm}[H]
\renewcommand{\thealgorithm}{1}
\caption{The K-means clustering algorithm}
\label{alg}
        \begin{algorithmic}[1]
                \State Randomly initialize $H$ centroids ${\bf{o}}_1,...,{\bf{o}}_H$
                \Repeat
                \For {$l \in \{1,...,L\}$ }
                \For{ $i \in \{1,...,N_l\}$ }
                \State Calculate cluster index $h_{i,l}$ for ${\bf{d}}_{i,l}$, i.e.,
                \begin{equation}
                h_{i,l} = \arg \min \limits_h dis({\bf{d}}_{i,l}, {\bf{o}}_h)
                \end{equation}
                \EndFor
                \EndFor
                \For {$h \in \{1,..H\}$}
                \State Recalculate the centroids
                \begin{equation}
                {\bf{o}}_h = \frac {{\sum\nolimits_{i,l}}{{\bf{1}}\{h_{i,l}=h\}}{\bf{d}}_{i,l}}{{\sum\nolimits_{i,l}}{{\bf{1}}\{h_{i,l}=h\}}}
                \end{equation}
                \EndFor
                \Until No shifts of centroids
        \end{algorithmic}
\end{algorithm}

However, the optimal number of the clusters $H$ is usually unknown. To obtain its value, clustering validation \cite{5694060} could be adopted. Specifically, we first use each candidate total cluster number to get different clustering results; then, the corresponding internal validation indices of each cluster is obtained, where the maximum candidate cluster number is set as $L$ in our formulation, which is the total number of areas; finally, we choose the best clustering result with the optimal cluster number according to the validation indices.

For the newly arriving UE in an area, it fetches the parameters from the parameter server controlling the cluster that includes this area, to have a jump start. Note that we need to jointly re-cluster all the UEs periodically based on the time scale of mobility pattern changes, which is usually on the order of hours \cite{Galati}. The details for handling the above aspects is skipped in this paper, in order to focus on other core issues.
\subsection{Asynchronous Joint Learning Scheme}
\begin{figure}[!htbp]
\setlength{\abovecaptionskip}{0pt}
\centering
\includegraphics [width=\linewidth]{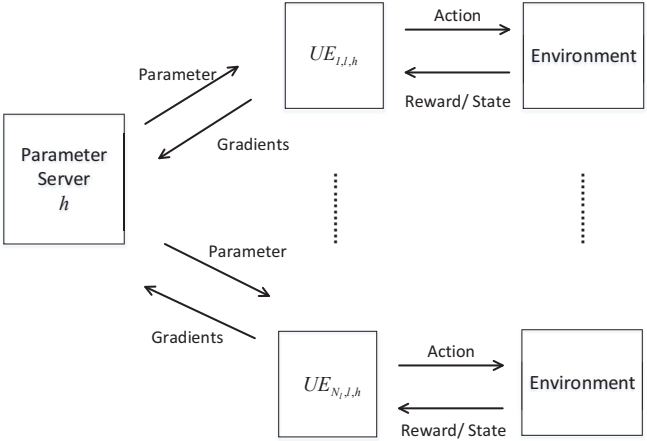}
\caption{ An example of asynchronous learning scheme in area $l$, where all the UEs are in cluster $h$. }
\label{fig3}
\end{figure}

After the clustering, the UEs are partitioned into clusters. In each cluster, an A3C framework is utilized to learn the optimal HO controller, as shown in Fig. \ref{fig3}.
If a UE requests to update the controller, it first fetches the most recent copy of the controller parameters from the parameter server, which stores the key global controller parameters. The UE then executes the controller and interacts with the environment, where the advantage actor-critic algorithm is used to compute the gradient of the controller network (assuming DQN is used). Finally, the UE pushes the new gradient back to the controller parameter server, which updates the global network parameter. Furthermore, the global controller weights in the parameter server are updated asynchronously in a lock-free fashion\cite{pmlr-v48-mniha16}. Note that the convergence property of this asynchronous learning framework has been established in [19] and we later show that this framework accelerates the individual HO controller learning as more UEs are joining the system.

\subsection{Action}
At time step $t$, the action $a_{i,l,h,t}$ for UE $i$ in area $l$ and cluster $h$ is a scalar representing the serving SBS index in ${\bf{b}}_{i,l,h,{t}}$.

\subsection{State Vector}
As mentioned above, the state vector represents the information for the situation and exploration. We use the RSRQs from all the SBSs to each UE representing the situation information and the serving SBS index to represent the exploration information. Hence, at time step $t$, the state vector for UE $i$ in area ${l}$ and cluster $h$ is given as
\begin{equation}
{\bf{s}}_{i,l,h,t}= \{ {\bf{q}}_{i,l,h,t}\;,\;{\bf{a}}_{i,l,h,{t-1}}\}
\end{equation}
where ${\bf{s}}_{i,l,h,t} \in {\bf{S}}_{i,l,h} $, ${\bf{q}}_{i,l,h,t} = \{q_{i,l,h,t}^1,...,q_{i,l,h,t}^{M_l}\}$ contains the RSRQs from all the candidate SBSs in ${\bf{b}}_{i,l,h,t}$ to UE $i$, and ${\bf{a}}_{i,l,h,{t-1}}$ is a one-hot encoded vector\footnote{The indices of SBSs are nominal labels, where the one-hot encoding is utilized to generate nominal vectors contained in state vectors.} according to the action $a_{i,l,h,{t-1}}$. In particular, a one-hot encoding \cite{silver2016mastering} is a representation of integral variables as binary vectors. Each integer value is represented as a binary vector that is all zero values except the index of the integer, which is marked with a one. For example, the action $a_{i,l,h,{t-1}}$ is an integer in the range of $[0,5]$ and its value $a_{i,l,h,{t-1}} = 2$. Accordingly, the one-hot vector ${\bf{a}}_{i,l,h,{t-1}}=[0,0,1,0,0,0]$.
\subsection{Reward Signal}
The reward signal should encourage the learning algorithm to achieve our goal, which is to ensure certain downlink throughput while minimizing the number of HOs. Here, the throughput is measured by the averaged throughput, which is defined as the ratio between the sum rate and the total time steps in the episode. Moreover, the HO performance is measured by the averaged HO rate, defined as the ratio between the HO times and the total time steps in the episode. According to the measures defined above, we set the reward signal as a weighted sum between the averaged throughput and the HO rate, which is available to the UE at the termination of each episode. The rewards for all the non-terminal steps could be set as zero. However, the resulting delayed rewards may cause the so-called credit assignment problem {\cite{712192}}, which degrades the performance of the RL approach.

To address this issue, we utilize reward shaping \cite{ng1999policy} to make the reward more informative and accelerate the training. In particular, at every time step $t$, the reward for UE $i$ in area $l$ and cluster $h$ is defined as
\begin{equation}
r_{i,l,h,t} = \left\{
\begin{aligned}
 R_{i,l,h,t}& - {\beta E,}&\text{if HO occurs}\\
&{ R_{i,l,h,t}}, &\text{otherwise}
\end{aligned}
\right.
\end{equation}
where $R_{i,l,h,t}$ is the rate of UE $i$ at time step $t$ in area $l$ and cluster $h$, $E$ is the energy consumption when a HO process occurs, and $\beta$ is a normalizing weight factor.
\section{Problem Formulation and Learning Algorithm}
In this section, we first formulate the RL problem for each UE after the clustering is done. The learning algorithm is then presented. Finally, we introduce the neural network structure of the controller and the novel idea of supervised-learning based network initialization.
\subsection{Problem Formulation and Learning Algorithm}
The policy for UE $i$ in area $l$ and cluster $h$ is defined as $\pi_{{\bm{\theta}}_{i,l,h}}$. Hence, the goal of UE $i$ in area $l$ and cluster $h$ is to obtain the optimal parameter ${\bm{\theta}}_{i,l,h}^*$ by solving the optimization problem (\ref{3}). For simplicity, we omit the subscripts $l$ and $h$ in the following derivations, unless explicitly mentioned. Hence, at time step $t$, the gradient to update parameter ${\bm{\theta}}_{i}$ is given as
\begin{equation}
\Delta {\bm{\theta}}_{i,t} ={{\nabla _{{\bm{\theta}}_{i}} }\log {\pi _{{\bm{\theta}}_{i}} }({a_{i,t}}|{\bf{s}}_{i,t})}Q^{\pi_{{\bm{\theta}}_{i}}}({\bf{s}}_{i,t},a_{i,t}).
\end{equation}

To estimate $Q^{\pi_{{\bm{\theta}}_{i}}}({\bf{s}}_{i,t+k},a_{i,t+k})$, we use a $n-$step prediction approach \cite{712192} combining the Monte-Carlo method \cite{sutton2000policy} and TD prediction \cite{712192} to balance the variance and bias caused by RL execution. In particular, the Monte-Carlo method alone obtains unbiased estimates but introduces a high variance, while TD learning alone could lower the variance but may introduce bias, especially when the estimation is not accurate. We here combine them: From time step $t$, UE $i$ interacts with the environment for the next $n$ time steps and obtains a trajectory $({\bf{s}}_{i,t},a_{i,t},r_{i,t},.....,{\bf{s}}_{i,t+n})$\footnote{If the terminal state occurs in the trajectory, the $n-$step prediction degrades to Monte-Carlo sampling without prediction.}.
We then use an estimator $V_{{\bf{w}}_{i}}({\bf{s}}_{i,t})$ parameterized by ${\bf{w}}_{i}$ to estimate the state-value function $V^{\pi_{{\bm{\theta}}_{i}}}({\bf{s}}_{i,t})$, with the estimation of $Q^{\pi_{{\bm{\theta}}_{i}}}({\bf{s}}_{i,{t}},a_{i,{t}})$ given as $Q_{{\bf{w}}_{i}}({\bf{s}}_{i,{t}},a_{i,{t}})= \sum\nolimits_{k' = t}^{n - 1} {{\gamma ^{k'-t}}{r_{i,k'}}}  + {\gamma ^{n-t}}{V_{{\bf{w}}_{i}}}({{\bf{s}}_{i
,t + n}})$. Consequently, the accumulated gradient from the sampled trajectory is given as
\begin{align}
&\Delta {\bm{\theta}}_{i}=\sum\limits_{k=0}^{n-1} \Delta {\bm{\theta}}_{i,t+k}\notag\\ &= \sum\limits_{k = 0}^{n-1}{{\nabla _{{\bm{\theta}}_{i}} }\log {\pi _{{\bm{\theta}}_{i}} }({a_{i,t+k}}|{{\bf{s}}_{i,t+k}})} A_{{\bf{w}}_{i}}({\bf{s}}_{i,t+k}),
\label{10}
\end{align}
where $\Delta {\bm{\theta}}_{i,t}$ is the gradient at time step $t$, $A_{{\bf{w}}_{i}}({\bf{s}}_{i,t+k})=Q_{{\bf{w}}_{i}}({\bf{s}}_{i,{t+k}},a_{i,{t+k}})-V_{{\bf{w}}_{i}}({\bf{s}}_{i,t+k})$ , and $V_{{\bf{w}}_{i}}({\bf{s}}_{i,t+k})$ is an inserted baseline \cite{712192} to further decrease the variance. In addition, the sampled gradient $\Delta {\bf{w}}_{i}$ is calculated as
\begin{align}
\Delta {\bf{w}}_{i}=\sum\limits_{k=0}^{n-1} \Delta {\bf{w}}_{i,t+k}= \sum\limits_{k = 0}^{n-1}{\nabla _{{\bf{w}}_{i}} }V_{{\bf{w}}_{i}}({\bf{s}}_{i,t+k})A_{{\bf{w}}_{i}}({\bf{s}}_{i,t+k}),
\label{11}
\end{align}
where $\Delta {\bf{w}}_{i,t}$ is the gradient at time step $t$, and $A_{{\bf{w}}_{i}}({\bf{s}}_{i,t})$ is estimated by forward-propagation in the approximator, which is independent of the back-propagation over ${\bf{w}}_{i}$.

Note that the gradient $\Delta {\bm{\theta}}_{i} $ is used to update the neural network weights in the policy improvement for $\pi_{{\bm{\theta}}_{i}}$ to approach the optimal policy incrementally. Meanwhile, ${Q}_{{\bf{w}}_{i}}({\bf{s}}_{i},a_{i})$ estimates $Q^{\pi_{{\bm{\theta}}_{i}}}({\bf{s}}_{i},a_{i})$ to evaluate the goodness of policy ${\pi_{{\bm{\theta}}_{i}}}$, such that (\ref{11}) is for the process of policy evaluation, i.e., value estimation. We see that (\ref{10}) is based on the derivative of the objective function, which is guaranteed to converge to a local optimum according to the stochastic approximation theorem \cite{712192}. To prove the convergence of policy evaluation following the gradient in (\ref{11}), we recall the Belleman operator $\Gamma ^{\pi_{{\bm{\theta}}_{i}}}: \left( {{\Gamma ^{\pi_{{\bm{\theta}}_{i}}} }} \right)Q({\bf{s}}_{i},a_{i})\buildrel \Delta \over = r_{i} + \left( P^{\pi_{{\bm{\theta}}_{i}}} \right)Q({\bf{s}}_{i},a_{i})$ \cite{NIPS2016}, where $Q({\bf{s}}_{i},a_{i})$ is an arbitrary $Q-$function for $({\bf{s}}_{i},a_{i})$, with ${\bf{Q}}^{\pi_{{\bm{\theta}}_{i}}}$\footnote{ ${\bf{Q}}^{\pi_{{\bm{\theta}}_{i}}}$ is the collection of the $Q-$fucntion $Q^{\pi_{{\bm{\theta}}_{i}}}({\bf{s}}_{i},a_{i})$ for all the state-action pairs.} defined as the unique fixed point of operator $\Gamma ^{\pi_{{\bm{\theta}}_{i}}}$, i.e., $\left(\Gamma ^{\pi_{{\bm{\theta}}_{i}}}\right){\bf{Q}}^{\pi_{{\bm{\theta}}_{i}}}={\bf{Q}}^{\pi_{{\bm{\theta}}_{i}}}$, where the operator $\Gamma ^{\pi_{{\bm{\theta}}_{i}}}$ is $\gamma-$contraction\cite{712192}, i.e., $\left\| {\left(\Gamma ^{\pi_{{\bm{\theta}}_{i}}}\right){\bf{Q}}_1-\left(\Gamma ^{\pi_{{\bm{\theta}}_{i}}}\right){\bf{Q}}_2} \right\|_\infty  \le \left\| {{\bf{Q}}_1-{\bf{Q}}_2} \right\|_\infty $ and $\left\| {{\bf{Q}}_1-{\bf{Q}}_2} \right\|_\infty = \max\nolimits_{({\bf{s}}_{i},a_{i})}|Q_1({\bf{s}}_{i},a_{i})-Q_2({\bf{s}}_{i},a_{i})|$. Note that the operator ${{P^{\pi_{{\bm{\theta}}_{i}}}}} $ is defined in (\ref{21}). Hence, from the contraction mapping theorem \cite{712192}, the iterative application of operator $\Gamma^{\pi_{{\bm{\theta}}_{i}}}$ to an arbitrarily $Q-$function converges to $Q^{\pi_{i}}$. When the $n-$step prediction target is sampled from $\left( {{\Gamma ^{\pi_{{\bm{\theta}}_{i}}} }} \right)^n Q_{{\bf{w}}_{i}}({\bf{s}}_{i},a_{i})$ with $\left( {{\Gamma ^{\pi_{{\bm{\theta}}_{i}}} }} \right)^n $ being $\gamma-$contraction, the policy evaluation process (\ref{11}) converges to $Q^{\pi_{{\bm{\theta}}_{i}}}$. Note that $\left( {{\Gamma ^{\pi_{{\bm{\theta}}_{i}}} }} \right)^n$ denotes $n$ successive applications of operator $ {{\Gamma ^{\pi_{{\bm{\theta}}_{i}}} }}$. In our algorithm, we implement the policy improvement and evaluation processes iteratively. The policy is improved with respect to the $Q-$function, while the $Q-$function is driven towards the true value function for the policy. It is easy to see that if both processes stabilize, then the estimated $Q-$function and policy are at least locally optimal \cite{712192}.

\begin{figure}[!htbp]
\setlength{\abovecaptionskip}{0pt}
\centering
\includegraphics [width=\linewidth]{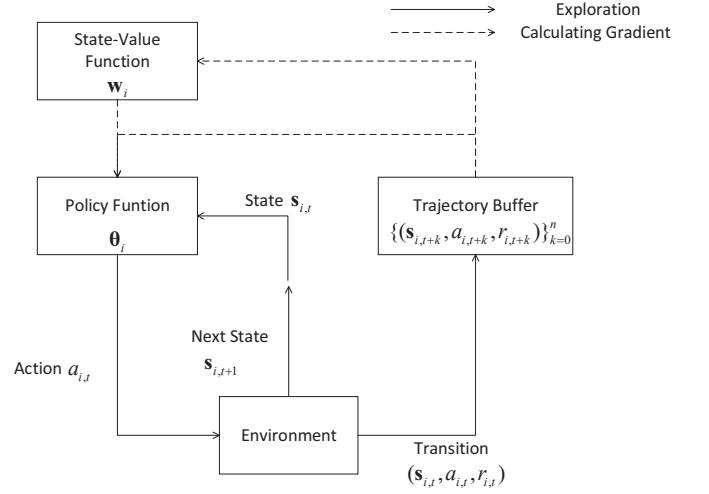}
\caption{ The overview of the learning algorithm implementation. }
\label{fig4}
\end{figure}
\subsection{Learning Algorithm Implementation}

When implementing the above learning algorithm, we need to follow the general RL framework as shown in Fig. \ref{fig1}. In particular, at time step $t$, the UE $i$ performs action $a_{i,t}$ according to the policy $\pi _{{\bm{\theta}}_{i}}$ given the state ${\bf{s}}_{i,{t}}$. Then, the environment generates the next state ${\bf{s}}_{i,{t+1}}$ and the reward signal $r_{i,t}$. The experience transition $({\bf{s}}_{i,{t}},a_{i,t},r_{i,t})$ is stored in the trajectory buffer; the UE receives the next state ${\bf{s}}_{i,{t+1}}$ and perform $a_{i,{t+1}}$ determined by $\pi _{{\bm{\theta}}_{i}}$, and this process continues. The UE accomplishes the interaction with $\pi _{{\bm{\theta}}_{i}}$ for $n$ time steps and the trajectory $({\bf{s}}_{i,t},a_{i,t},r_{i,t},.....,{\bf{s}}_{i,t+n})$ is stored in the buffer. Note that we could obtain ${V_{{\bf{w}}_{i}}}({{\bf{s}}_{i,t + n}})$ if ${\bf{s}}_{i,t+n}$
is given. Moreover, the state-value function at each step in the sampled trajectory can be estimated by forward-propagation in the approximator. Hence, for each state ${\bf{s}}_{i,t+k}, k \in \{0,1,...,n\}$ in the trajectory, we obtain $A_{{\bf{w}}_{i}}({\bf{s}}_{i,t+k})$, which is used to calculate the accumulated gradients $\Delta {\bm{\theta}}_{i}$ and $\Delta {\bf{w}}_{i}$ according to (\ref{10}) and (\ref{11}), respectively, by back-propagation in the neural network approximator. The whole framework is shown in Fig. \ref{fig4}.
\begin{figure}[!htbp]
\setlength{\abovecaptionskip}{0pt}
\centering
\includegraphics [width=\linewidth]{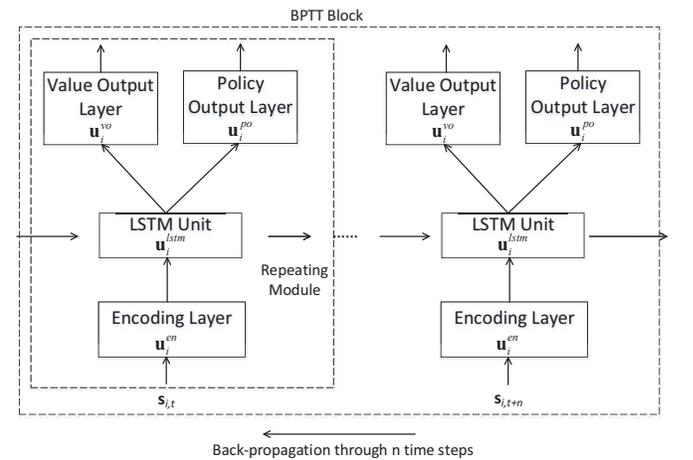}
\caption{ The structure of DQN for state value estimation and policy generations. }
\label{fig5}
\end{figure}
\subsection{DQN in HO Controller}

As shown in Fig. \ref{fig5}, for UE $i$ in area $l$ and cluster $h$, we utilize a three-layer neural network to generate $\pi_{{\bm{\theta}}_{i}}$ and estimate $V_{{\bf{w}}_{i}}$, including the encoding layer, the LSTM \cite{Sutskever} layer and the output layer, where the encoding neuron layer and the output layer are fully connected and the LSTM layer consists of multiple LSTM units as we discussed earlier. In particular, at time step $t$, a repeating module includes the encoding layer, the LSTM unit, and the value and policy output layers with weights ${\bf{u}}_{i,t}^{en}$, ${\bf{u}}_{i,t}^{lstm}$, ${\bf{u}}_{i,t}^{vo}$, and ${\bf{u}}_{i,t}^{po}$, respectively. Note that the weights in the repeating modules are shared across all the time steps in the same back-propagation truncation block, which will be clarified later. Accordingly, the parameters for the actors and critics are ${{\bm{\theta}}_{i}}=[{\bf{u}}_{i}^{en},{\bf{u}}_{i}^{lstm},{\bf{u}}_{i}^{po}]$ and ${{\bf{w}}_{i}}=[{\bf{u}}_{i}^{en},{\bf{u}}_{i}^{lstm},{\bf{u}}_{i}^{vo}]$, respectively, with subscript $t$ neglected. We see that the actors and critics share the weights of the encoding and LSTM layers, and the weights of the whole network could be denoted as ${{\bf{u}}_{i}}=[{\bf{u}}_{i}^{en},{\bf{u}}_{i}^{lstm},{\bf{u}}_{i}^{vo},{\bf{u}}_{i}^{po}]$. Furthermore, the weights ${{\bm{\theta}}_{i}}$ and ${{\bf{w}}_{i}}$ for UE $i$ in cluster $h$ are the copies of the global weights ${{\bm{\theta}}_h}=[{\bf{u}}_{h}^{en},{\bf{u}}_{h}^{lstm},{\bf{u}}_{h}^{po}]$ and ${{\bf{w}}_h}=[{\bf{u}}_{h}^{en},{\bf{u}}_{h}^{lstm},{\bf{u}}_{h}^{vo}]$, which are stored and updated in the parameter servers serving the UEs in cluster $h$.

\begin{figure}[!htbp]
\setlength{\abovecaptionskip}{0pt}
\centering
\includegraphics [width=\linewidth]{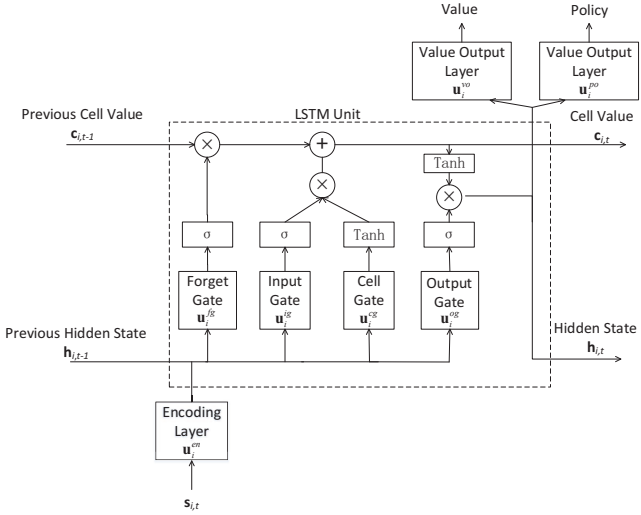}
\caption{ The structure of repeating module. }
\label{fig10}
\end{figure}
The structure of the repeating module is shown in Fig. \ref{fig10}.
Specifically, the LSTM unit can explore a self-learned amount of long-range temporal information, and it consists of the forget, input, cell and output gates with weights of ${\bf{u}}_{i}^{fg}$, ${\bf{u}}_{i}^{ig}$, ${\bf{u}}_{i}^{cg}$ and ${\bf{u}}_{i}^{og}$, respectively. In the forward-propagation, the cell value ${\bf{c}}_{i,t}$ and the hidden state ${\bf{h}}_{i,t}$ at time step $t$ are given as
\begin{align}
&{{\bf{c}}_{i,t}} = {{\bf{c}}_{i,t - 1}} \times {\bf{f}}_{i}^{fg} + {\bf{f}}_{i}^{ig} \times {\bf{f}}_{i}^{cg},\\
&{{\bf{h}}_{i,t}} = {\bf{f}}_{i}^{og} \times \tanh ({{\bf{c}}_{i,t}}),
\end{align}
where ${\bf{f}}_{i}^{fg}$, ${\bf{f}}_{i}^{ig}$, ${\bf{f}}_{i}^{cg}$ and ${\bf{f}}_{i}^{og}$ are the outputs of forget, input, cell and output gates defined as\footnote{In this paper, we drop the bias for simplicity.}
\begin{align}
&{\bf{f}}_{i}^{fg} = \sigma ({\bf{u}}_{i}^{fg} \times [{\bf{f}}_{i}^{en} ,{{\bf{h}}_{i,t - 1}}]),\\
&{\bf{f}}_{i}^{ig} = \sigma ({\bf{u}}_{i}^{ig} \times [{\bf{f}}_{i}^{en} ,{{\bf{h}}_{i,t - 1}}] ),\\
&{\bf{f}}_{i}^{cg} = \tanh ({\bf{u}}_{i}^{cg} \times [{\bf{f}}_{i}^{en},{{\bf{h}}_{i,t - 1}}] ),\\
&{\bf{f}}_{i}^{og} = \sigma ({\bf{u}}_{i}^{cg} \times [{\bf{f}}_{i}^{en} ,{{\bf{h}}_{i,t - 1}}] ),
\end{align}
 respectively, with ${\bf{f}}_{i}^{en}={\bf{u}}_{i}^{en} \times {{\bf{s}}_{i,t}}$ being the output of encoding layer. Moreover, we have the activation functions as $\sigma(x) = \frac{1}{1+e^{-x}}$ and $tanh(x) = \frac{e^{x}-e^{-x}}{e^{x}+e^{-x}}$, and ${\bf{u}}_{i}^{lstm} = [{\bf{u}}_{i}^{fg}, {\bf{u}}_{i}^{ig}, {\bf{u}}_{i}^{cg}, {\bf{u}}_{i}^{og}]$. Furthermore, the outputs of the actors are vectors (the policies) with dimension $M_l$ to represent the probability distribution of choosing among the $M_{l}$ SBSs, which is ${{\pi}}_{{\bm{\theta}}_{i}}(a_{i,t}|{\bf{s}}_{i,t})=softmax({\bf{u}}_{i,t}^{po}\times{{\bf{h}}_{i,t}})$ at time step $t$, with $softmax({\bf{x}}) = [\frac{e^{x_1}}{\sum\nolimits_{k=1}^{|{\bf{x}}|}e^{x_k}},...,\frac{e^{x_{|{\bf{x}}|}}}{\sum\nolimits_{k=1}^{|{\bf{x}}|}e^{x_k}}]$. In addition, the action $a_{i,t}$ is chosen according to the policy ${{\pi}}_{{\bm{\theta}}_{i}}$. The outputs of the critics are scalars to estimate the state-value function, which is $V_{{\bf{w}}_{i}}({\bf{s}}_{i,t})={\bf{u}}_{i,t}^{vo}\times{{\bf{h}}_{i,t}}$ at time step $t$.
For a communication system, utilizing LSTM has two advantages: The LSTM units can capture the moving patterns from the historical location information; and it can also accumulate the exploration information, which leads to better tradeoff between exploration and exploitation.

Furthermore, we apply the truncated back-propagation through time (BPTT) technique \cite{Sutskever} to alleviate error accumulation, where BPTT is executed every $n$ steps. The initial cell state and hidden state for the BPTT block are from the cell and hidden state outputs of the LSTM unit at the last time step in the previous BPTT block\footnote{If the terminal state occurs in the previous BPTT block, the initial cell and hidden states are both set as zero vectors for the current BPTT block.}. In a BPTT block, the gradient with respect to ${\bf{u}}_{i}$ at the time step $t$ is given as
\begin{align}
&\Delta {\bf{u}}_{i,t}= \Delta {\bm{\theta}}_{i,t}+\Delta {\bf{w}}_{i,t}\notag\\
&={\nabla _{{\bf{u}}_{i}} } [\log {\pi _{\bm{\theta}_{i}} }({a_{i,t+k}}|{{\bf{s}}_{i,t+k}})+V_{{\bf{w}}_{i}}({\bf{s}}_{i,t})] A_{{\bf{w}}_{i}}({\bf{s}}_{i,t}).
\label{16}
\end{align}
After obtaining the gradients at each time step in the BPTT block from (\ref{16}), the accumulated gradients $ \Delta {\bm{\theta}}_{i}$ and $\Delta {\bf{w}}_{i}$ are derived according to (\ref{10}) and (\ref{11}).

Note that we have adopted the asynchronous RMSProp \cite{pmlr-v48-mniha16} routine to update the parameters in cluster $h$, which is given as
\begin{align}
{\bf{g}}_{h} = \alpha^\prime {\bf{g}}_{h}+ (1 - \alpha^\prime )\Delta {\bm{\theta}_{i}} ^2
\label{12}
\end{align}
where $\alpha^\prime$ is a decay factor. In asynchronous RMSProp, ${\bf{g}}_{h}$ is used to update the DQN weights $\bm{\theta}_h$ in cluster $h$ as
\begin{align}
\bm{\theta}_h  \leftarrow \bm{\theta}_h  - \eta \frac{\Delta {\bm{\theta}_{i}}}{{\sqrt {{\bf{g}}_{h} + \varepsilon } }},\label{13}
\end{align}
with $\varepsilon$ a small constant to prevent the dominator from approaching 0, and ${\bm{\theta}}_h$ sharable to all the UEs in cluster $h$. Note that the multiplication, the division, the square and the square root operations in (\ref{12}) and (\ref{13}) are all element-wise. The way for handling the global update for ${\bf{w}}_h$ with $\Delta {\bf{w}}_{i}$ is the same as those in (\ref{12}) and (\ref{13}). The full algorithm for UE $i$ in area $l$ and cluster $h$ to update HO controller in cluster $h$ is given in Algorithm \ref{alg2}, where we assume that $t$ is the local time step counter for each UE, $t_{global}$ is the global time step counter for all the UEs, $t_{start}$ is the local record of the starting time step for each UE.
\begin{algorithm}
\caption{The asynchronous advantage actor-critic learning algorithm for UE $i$ in area $l$ and cluster $h$}
\label{alg2}
        \begin{algorithmic}[1]
                \State Initialize the local counter as $t=1$
                \Repeat \State Reset accumulated gradients: $\Delta {\bm{\theta}}_{i} = \Delta {\bf{w}}_{i} =0$;
                \State Fetch parameters ${\bm{\theta}}_{i}$ and ${\bf{w}}_{i}$ from parameter server, i.e., copying ${\bm{\theta}}_h$ and ${\bf{w}}_h$;
                \State Set $t_{start}=t$ and get initial state ${\bf{s}}_t$
                \Repeat \State Perform ${a_{i,t}}$ according to the policy ${\pi _{{\bm{\theta}}_{i}} }({a _{i,t}}|{{\bf{s}}_{i,t}})$
                \State Set $t \leftarrow t+1$ and $t_{global} \leftarrow t_{global}+1$
                \Until terminal ${\bf{s}}_{i,t}$ or $t-t_{start}==n$
                \begin{equation}
                r = \left\{
                \begin{aligned}
                &0,     &\text{for terminal ${\bf{s}}_{i,T}$},\notag\\
                &V_{{\bf{w}}_{i}}({\bf{s}}_{i,t}), &\text{otherwise},\notag
                \end{aligned}
                \right.
                \end{equation}
                \For {$j \in \{t-1,...,t_{start}\}$ }
                \State $r \leftarrow r_{i,j}+\gamma r$
                \State Accumulate the gradients, i.e.,
                \begin{align}
                &\Delta {\bm{\theta}}_{i} \leftarrow \Delta {\bm{\theta}}_{i} +
                 (r -V_{{\bf{w}}_{i}}({\bf{s}}_{i,j})){\nabla _{{\bm{\theta}}_{i}} }\log {\pi _{{\bm{\theta}}_{i}} }({a_{i,j}}|{{\bf{s}}_{i,j}}) \notag\\&+ \eta \sum\limits_a \pi_{{\bm{\theta}}_{i}}(a|{{\bf{s}}_{i,j}})log\pi_{{\bm{\theta}}_{i}}(a|{{\bf{s}}_{i,j}}) \\
                 &\Delta {\bf{w}}_{i} \leftarrow \Delta {\bf{w}}_{i}+ (r -V_{{\bf{w}}_{i}}({\bf{s}}_{i,j})){\nabla _{{\bf{w}}_{i}} }V_{{\bf{w}}_{i}}({\bf{s}}_{i,j})
                \end{align}
                \EndFor
                \State Using (\ref{12}) and (\ref{13}) to update ${\bm{\theta}}_h$ and ${\bf{w}}_h$ in parameter server with accumulated gradients $\Delta {\bm{\theta}}_{i}$ and $\Delta {\bf{w}}_{i} $
                \Until the learning is done.
        \end{algorithmic}
\end{algorithm}
\subsection{Initialization with Supervised Learning }
To boost the performance of the RL framework and fully utilize what we have already known from the traditional model-based HO strategies, we further propose to initialize the involved neural networks with supervised learning. In particular, we assume that UE $i$ in area $l$ and cluster $h$ receives the RSRPs in its active set ${\bf{b}}_{i,t}$ at time step $t$. If one of the SBSs in ${\bf{b}}_{i,t}$ satisfies the following condition for a
specific time determined by the TTT:
\begin{equation}
RSRP_{b_{i,t} \in {\bf{b}}_{i,t}} > RSRP_{b_{i,t}^{serving}} + HHM,
\end{equation}
where $RSRP_{b_{i,t}^{serving}}$ is the RSRP of the serving SBS for UE $i$ in area $l$ and cluster $h$ at time step $t$, the UE hands itself over to the SBS with index $b_{i,t}$. Hence, we obtain the action $a_{i,t}$. Moreover, we obtain ${\bf{s}}_{i,t}$ over the RSRPs from all the SBSs in the ${\bf{b}}_{i,t}$ and the serving SBS index. To generate the training data, we incorporate a similar technique to that in \cite{Boje2e}, where we use five different pairs of HHMs and TTTs. Specifically, in cluster $h$, we sample five episodes as training data, i.e., $\{({\bf{s}}_{i,1}^j,a_{i,1}^j),...,({\bf{s}}_{i,T}^j,a_{i,T}^j)\}_{j=1}^5$.
Note that we can only initialize the encoding layer ${\bf{u}}_{h}^{en}$, LSTM layer ${\bf{u}}_{h}^{lstm}$, and policy softmax output layer ${\bf{u}}_{h}^{po}$, while the value output layer ${\bf{u}}_{h}^{vo}$ is initialized randomly, due to the absence of knowledge on system dynamics.

\section{Simulation}
In this section, the simulations setups are shown firstly. Then the simulation results are presented. Finally, the upper bound of number of UEs in each cluster is discussed.
\subsection{Simulation Setups}

In the experiment, we have $L=3$ nonadjacent areas, i.e., $\mathscr{A}_1$, $\mathscr{A}_2$ and $\mathscr{A}_3$. Each area has 6 SBSs, i.e., $K_1 = K_2 = K_3 = 6$. We assume that all the three areas are $16 \times 16 $ square meters and the SBSs are deployed in each area randomly. Moreover, a random walk model \cite{WCM72} is utilized to simulate the movements of UEs. In particular, at every time step, the UEs move in four directions, i.e., east, south, west and north. In addition, the probability distribution of the four directions are set as $[0.25,0.25,0.25,0.25]$ in $\mathscr{A}_1$ and $\mathscr{A}_2$, and $[0.6,0.2,0.1,0.1]$ in $\mathscr{A}_3$. The speed levels of UEs are chosen randomly from interval $[1,3]$ $m/s$. The transmit power of all SBS is set as $30 dBm$, with bandwidth $10 MHz$ and thermal noise density $-174 dBm/Hz$. The interruption time of HO is up to $50ms$ \cite{RRC}, and the energy consumption for the HO is $E=0.3$ $Joule$. The pathloss is given as $PL(dB) = 36.7\log_{10}(distance)+39.4$\cite{7987783} and the log-normal shadowfading is with zero mean and standard deviation $8 dB$. The small-scale fading is assumed to be Rayleigh distributed. The length of the BPTT block is set as $n=20$. The encoding, policy output and value output layers are fully connected with sizes $12\times8$, $8\times6$ and $8\times1$, respectively. We further set $\eta= 0.01$ for entropy regularization and $\alpha ^\prime = 0.99$ for the RMSProp decay factor. Finally, we set the time period for mobility observation in of clustering is $T_u = 150$ time slots. Note that the interferences are not considered in our paper.

\subsection{Simulation Results}
First of all, the clustering validation result is shown in Table \ref{table1}, where we utilize the Calinski-Harabaz index (CHI) \cite{5694060} as the measure to evaluate the goodness of clustering with different cluster numbers. In particular, the CHI is defined as:
\begin{equation}
{\rm{CHI}}(k) = \frac{{\sum\limits_{h = 1}^H {N_h dis({ {{{\bf{o}}_h},{\bf{o}}} })} }}{{\sum\limits_{h = 1}^H {\sum\limits_{{{\bf{G}}_h}} dis({{ {{{\bf{d}}_{i,l,h}},  {{\bf{o}}_h}} }}) } }} \times \frac{{\left| {\bf{D}} \right| - H}}{{H - 1}},
\label{15}
\end{equation}
where ${\bf{o}}$ is the centroid of the whole data set ${\bf{D}}$ and $|{\bf{D}}|$ is the size of the data set. Note that the numerator and the dominator in (\ref{15}) represent the inter-cluster separation and the intra-cluster compactness, respectively. Hence, the cluster number with the largest CHI is the optimal cluster number. Table \ref{table1} shows the CHI values with different data set sizes, where we assume that each area has 4, 40 and 400 UEs, respectively. Moreover, the CHI values are averaged over 100 random runs. From the result, we see that it is better to partition the UEs into two clusters under the current setup. In the following, we assume that each area has 4 UEs for simplicity.
\begin{table}[H]
\centering
\caption{Clustering validation results with Calinski-Harabaz Index}
\label{table1}
\resizebox{0.9\linewidth}{!}{%
\begin{tabular}{|c|c|c|ll}
\cline{1-3}
\multicolumn{1}{|l|}{\multirow{2}{*}{Total UEs}} & \multicolumn{2}{l|}{Calinski-Harabaz Index} &  &  \\ \cline{2-3}
\multicolumn{1}{|l|}{} & H=2 & H=3 &  &  \\ \cline{1-3}
12 & 31.97 & 26.09 &  &  \\ \cline{1-3}
120 & 300.99 & 208.50 &  &  \\ \cline{1-3}
1200  & 2964.97 & 1986.55 & & \\ \cline{1-3}
\end{tabular}%
}
\end{table}

\begin{figure}[!htbp]
\setlength{\abovecaptionskip}{0pt}
\centering
\includegraphics [width=\linewidth]{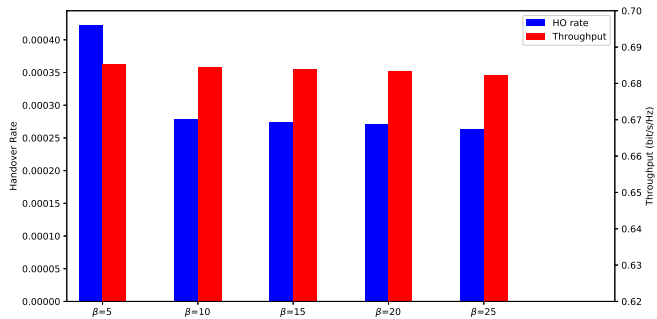}
\caption{ The tradeoff between HO rate and throughput when the weight factor $\beta=5,10,15,20,25$.  }
\label{fig8}
\end{figure}

Then, in Fig. \ref{fig8}, the tradeoff between the HO rate and throughput is shown by changing the weight factor $\beta$ in the reward signal. We see that the rewards with larger $\beta$ could encourage the controller to lower the HO rate by sacrificing the throughput.

\begin{figure}[!htbp]
\setlength{\abovecaptionskip}{0pt}
\centering
\includegraphics [width=0.94\linewidth]{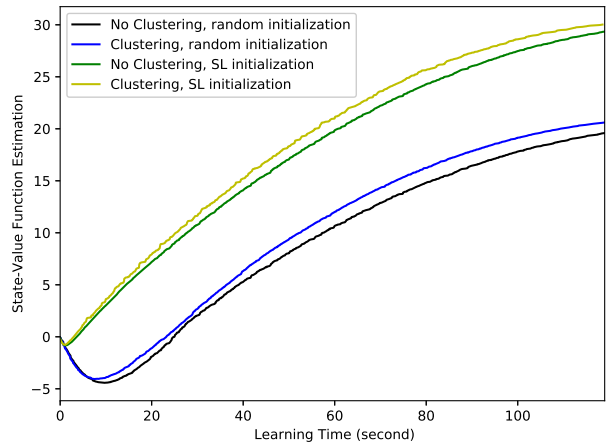}
\caption{ The estimation of the state-value function versus learning time.}
\label{fig6}
\end{figure}

After the clustering, the UEs in areas $\mathscr{A}_1$ and $\mathscr{A}_2$ are partitioned into the same cluster, with the others into the second cluster. In Fig. \ref{fig6}, we show the estimation of the state-value function with/without SL-based initialization for the UEs in area $\mathscr{A}_1$, where the total learning time is 119 seconds, and $\beta=25$.  In particular, the learning without SL-based initialization starts with randomly initialized weights. According to the definition of state-value function, a larger state value means that the current policy gains more expected rewards in the future. Hence, SL-based initialization could help the A3C framework derive a better policy through the same length of learning period. It is worth noting that the state value may decay at the beginning, for the policy is stochastic and the exploration in the parameter space is based on the current policy. Hence, if the initial policy is bad, the exploration can cause certain performance degradation. We observe that SL-based initialization network could mitigate such degradation, since the 3GPP HO policy used in SL is typically better than a random policy. To illustrate the gain of clustering, where we could group more UEs with similar mobility patterns into one cluster such that the shared global parameter server could exploit more information, we also draw the case without clustering, i.e., only the UEs in $\mathscr{A}_1$ share one parameter server instead of all the UEs in $\mathscr{A}_1$ and $\mathscr{A}_2$ share one parameter server. The results in Fig. \ref{fig6} show that the proposed asynchronous RL framework could achieve larger state values with clustering. It is due to the fact that clustering with a global parameter server per cluster allows more UEs to update the weights in the shared parameter server, which could exploit more sampled data in the same time period. Hence, the A3C framework accelerates learning when the number of UEs increases in the cluster, which is a strong indicator that the proposed method could be applied to large systems.

For the UCB learning algorithm \cite{7572176,7925950} in comparison, the action and reward are set the same as those in our work. The action for UE $i$ in area $l$ and cluster $h$ at time step $t$ is given as
\begin{align}
{a_{i,l,h,t}} = \arg \mathop {\max }\limits_{k = 1,...K} \left( {{\mu _{i,l,h,k}} + \sqrt {\frac{{2\ln t}}{{{M_{i,l,h,k,t - 1}}}}} } \right)\\
{\mu _{{a_{i,l,h,t}}}} \leftarrow {\mu _{{a_{i,l,h,t}}}} + \frac{1}{{{M_{i,l,h,{a_{i,t}},t}}}}({r_{i,l,h,t}} - {\mu _{i,l,h,{a_{i,l,h,t}}}}) \label{14}
\end{align}
where ${\mu _{i,l,h,k}}$ is the estimated mean of the reward distribution of SBS $k$ for UE $i$ in area $l$ and cluster $h$, which is updated as in (\ref{14})\footnote{At time step $t$, $k=a_{i,l,h,t}$.}, and $M_{i,l,h,k,t - 1}$ is the number of times that UE $i$ selects SBS $k$ up to time step $t$.
\begin{figure}[!htbp]
\setlength{\abovecaptionskip}{0pt}
\centering
\includegraphics [width=\linewidth]{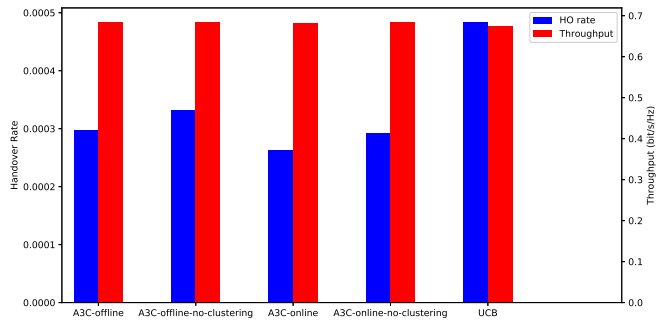}
\caption{ The averaged throughput and HO rate in the testing time. }
\label{fig7}
\end{figure}

In Fig. \ref{fig7}, we compare the averaged HO rates and throughputs of UCB, SL initialized A3C-online and A3C-offline RL methods with/without clustering for a newly arriving UE. Note that we randomly pick one learned network from the 100 realizations used in Fig. \ref{fig6} and consider it as the pretrained network for testing, where the testing period is 2 seconds. In particular, the pretrained network is used to control the UE HOs without further learning in the off-line method. In the on-line method, the UE keeps learning with the same A3C framework. Moreover, the newly arriving UE with UCB needs to learn with random initialization. Hence, for fairness, the HO rate and throughput for UCB are averaged over 121 seconds including the learning and testing periods. Similar to the on-line method, the UE with UCB keeps learning during testing. Furthermore, such generated HO rates and throughputs are all averaged over 500 random runs. We see that our proposed methods could gain better throughputs and lower HO rates after equal learning time compared with the UCB algorithm, which is already up to $80\%$ better than the 3GPP method \cite{7925950}. In addition, the clustering with a global parameter server per cluster further improves the performance.

\subsection{Discussion}
The asynchronous learning scheme suffers from a critical issue, which is called delayed-gradient. Specifically, before UE $i$ in area $l$ and cluster $h$ wants to push the gradient $\Delta {\boldsymbol{\theta}_{i,l,h}}$ (calculated based on the global weights ${\boldsymbol{\theta}_{h}}$ in cluster $h$ at a global time step $t_{global}$) to the parameter server of cluster $h$, several other UEs may have already pushed in their gradients, and the global weights ${\boldsymbol{\theta}_{h}}$ in the parameter server may have been updated for $\Delta t_{global}$ time steps, where $\Delta t_{global}$ is called the delay factor. Moreover, the upper bound of the delay factor $\Delta t_{global}$ is roughly proportional to the number of UEs. According to the recent results \cite{Lian15}, the number of UEs should be upper-bounded by $O(\sqrt{t_{ps}^h})$ \cite{Lian15} to ensure that the asynchronous learning scheme can accelerate the learning.
\section{Conclusions}

In this study, we proposed a two-layer framework to optimize the HO processes in the goal of lowering the HO rates and ensuring the system throughputs, where the user mobility patterns could be heterogeneous. In the proposed framework, a centralized controller first partitioned the UEs with different mobility patterns into clusters. Then, an asynchronous learning scheme is adopted in each cluster. In particular, we let the UEs fetch the most recent copy of DQN approximators from a global parameter server in each cluster, then let them execute the policy and compute the gradients using the observed states. Afterwards, the new gradients are pushed back to the parameter server for updating the global parameter. Thanks to the generalization ability of DNN, newly arriving UEs can use the pretrained neural networks from the parameter server to avoid the likely ill-performed initial points. To further improve the performance, we use SL based on data from traditional HO methods to initialize the DQN approximators before the execution of RL, to compensate the negative effects caused by exploration in the early stages of learning.

\IEEEpeerreviewmaketitle

\ifCLASSOPTIONcaptionsoff
  \newpage
\fi

\bibliographystyle{IEEEtran}
\bibliography{quotation}
\end{document}